# High Sensitivity Collimators for Optimising Lesion Detection in SPECT Images

Krzysztof Kacperski [a), b)], Dominika Świtlik [a), c)], and Jakub Pietrzak [a), c)]

*Abstract–* Parallel hole collimators which are currently routinely used in SPECT imaging were designed a few decades ago, when filtered backprojection (FBP) was used for tomographic reconstruction. Statistical reconstruction methods with precise modeling of the projection measurement offer a different standard, and the question of optimizing resolution – sensitivity tradeoff by choosing appropriate collimator aperture needs to be revised. In this paper we search for a parallel hole collimator which offers best performance in detection of hot lesions on a uniform background. To evaluate the image quality we use standard quantitative measures like the contrast recovery coefficient, coefficient of variation, and contrast to noise ratio. We also performed signal detection by human observers followed by Receiver Operating Characteristics (ROC) analysis for a few lesions close to the limit of visibility. Our results consistently indicate that optimal performance is achieved with collimators of sensitivity 6 – 9 times that of the conventional high resolution (HR) collimator. When the optimized collimator is applied, the image quality of the conventional image obtained with the HR can be achieved at scanning time or activity dose reduced by factor of 3.

## I. INTRODUCTION

COLLIMATOR is one of the key elements of a gamma camera, controlling the resolution-noise (counts statistics) trade-off inherent to nuclear medicine imaging. By opening up the aperture more photons can be collected, however they carry less information on the original activity distribution; the projected image is more blurred. The choice of a collimator which provides optimal compromise is not trivial [1], particularly when nonlinear iterative reconstruction methods are used. In general it depends on the imaged object and the task which is to be solved by examining the image. The problem has been studied by many researchers using different tasks and optimisation criteria [2-7].

In this study we address the question: which parallel hole collimator offers optimal resolution-noise compromise with respect to hot lesion detectability, when the statistical reconstruction methods with collimator resolution modeling are used. We simulate multiple hot lesions of variable size and contrast on a uniform low activity background scanned with a range of collimators and exposition times.

## II. MATERIALS AND METHODS

### A. Phantoms

We used two cylindrical numerical phantoms of the diameter 42 cm filled uniformly with water solution of 200 MBq $^{99m}$Tc to simulate the background. 12 spherical hot lesions have been placed symmetrically at the distance 12.5 cm from the central axis of the phantom in three layers.

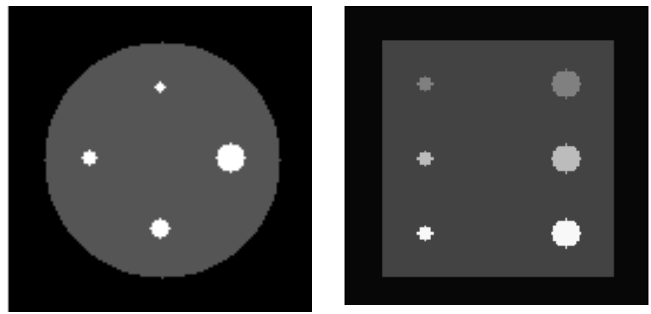

Fig. 1. Numerical Phantom with "big" hot lesions: transaxial (left) and lateral (right) cross-sections.

One phantom contained small lesions of diameters 10, 14, 16 and 17 mm and tumour to background contrast ratios 5:1, 10:1 and 15:1 for each lesion size. The other phantom contained big lesions of diameters 17, 26, 33 and 50 mm of lower contrast ratios: 2:1, 3:1 and 4:1 respectively at the same positions.

### B. Scanning and reconstruction; collimators

SPECT scans with 120 projection angles over 360° have been simulated and the images reconstructed using the Ordered Subsets Expectation Maximisation (OSEM) algorithm with 5 subsets to avoid excessive noise. Photon attenuation and collimator/detector point response function have been modelled both in projection and the reconstruction; Compton scatter has been neglected. Inter-iteration Gaussian smoothing has been used to control noise in the reconstruction. For each noise-free projection 30 instances of Poisson noise were generated to assess image noise. The count levels in projections were scaled to simulate different scanning times for a single detector head. The results below are shown for smoothing kernel with $\sigma = 2.5$ mm.

We tested a range of parallel hole collimators, mainly of high sensitivity; some of them are listed in the table below.

This work was supported by the Wellcome Trust under Grant No. 084288/Z/07/A.
a) The Maria Skłodowska-Curie Memorial Cancer Centre and Institute of Oncology, Warsaw, Poland.
b) National Centre for Nuclear Research, Świerk, Poland
c) Faculty of Physics, University of Warsaw, Poland..

email: Krzysztof.Kacperski@ncbj.gov.pl

TABLE I. COLLIMATORS USED IN THE STUDY

| Collimator type | Symbol | FWHM [mm] at 10 cm | Relative sensitivity |
|---|---|---|---|
| High resolution | HR | 7.1 | 1 |
| High sensitivity | HS6 | 12.5 | 5.94 |
|  | HS9 | 15.4 | 9.28 |
|  | HS14 | 18.5 | 14.06 |

### C. Image quality measures

To quantify image noise 50 background masks of the same sizes as the lesions were created and placed randomly in 5 layers, 10 masks in each: three layers with the lesions and additional two symmetrically between them.

The following conventional parameters were computed for each set of scan and reconstruction parameters (cf. [8]):

Contrast:
$$C_l = \left\langle \frac{M_l - \overline{M_{bgd}}}{\overline{M_{bgd}}} \right\rangle$$

where $M_l$ is the average number of counts in the lesion, and $M_{bgd}$ – the average number of counts in the background masks. $\overline{M}$ denotes the average over the background masks within the layer of the lesion of a single image, and $\langle M \rangle$ is the ensemble average over the 30 noisy image instances.

Contrast recovery coefficient (CRC):
$$CRC = \frac{C_{rec}}{C_{org}}$$

average Coefficient of Variation (COV) in the background region:
$$COV = \frac{\overline{\sqrt{var\langle M_{bgd}\rangle)}}}{\langle M_{bgd} \rangle}$$

where $\bar{x}$ is the average over all the background masks.

Contrast to Noise Ratio (CNR):
$$CNR_l = \frac{\langle M_l - M_{bgd} \rangle}{\sqrt{\frac{1}{2}(var\langle M_l\rangle + var\langle M_{bgd}\rangle)}}$$

where $M_{bgd}$ is the average number of counts in a single background mask located below or above the respective lesion. To obtain an overall measure of image quality with respect to all the lesions we introduced an effective number of visible lesions in the phantom defined as
$$N_L = \sum_{l=1}^{12} g(CNR_l)$$

where $g(x) = 1 - (1/(1 + (x/\alpha)^k))$ is a sigmoidal function of the $CNR$. We used the parameter values $\alpha = 6, k = 5$. For the lesions clearly visible (large CNR) $g(CNR) \approx 1$; for invisible lesions $g(CNR) \approx 0$.

We also performed human observer detection tests for a few lesions (SKE/BKE type task), followed by the ROC analysis of the obtained results. Two (the same) observers evaluated a series of 60 images for each lesion, divided equally between positive and negative cases, after going through a learning set of the same size. 5-point categorical rating scale was used to evaluate the level of confidence in the outcome. The mean values of the areas under ROC curves (AUC) obtained by both observers are presented as the final result.

## III. RESULTS

Without regularization the reconstructed images show excessive noise and are not useful for lesion detection. We use Gaussian smoothing of the image estimate before each next iteration to control noise. Smoothing kernel with $\sigma = 2.5$ mm yields close to optimal performance in terms of the maximal CNR for most lesions. The reconstruction algorithm converges to approximately stationary solution after about 50 OSEM (ML_EM equivalent) iterations.

Fig. 1 shows examples of images obtained with different collimators. The conventional HR provides high lesion contrast, but also high background noise. The images obtained with a high sensitivity collimator exhibit a pronounced blur of the lesions (partial volume effect), however due to the reduced background noise, more small/low contrast ones can actually be detected.

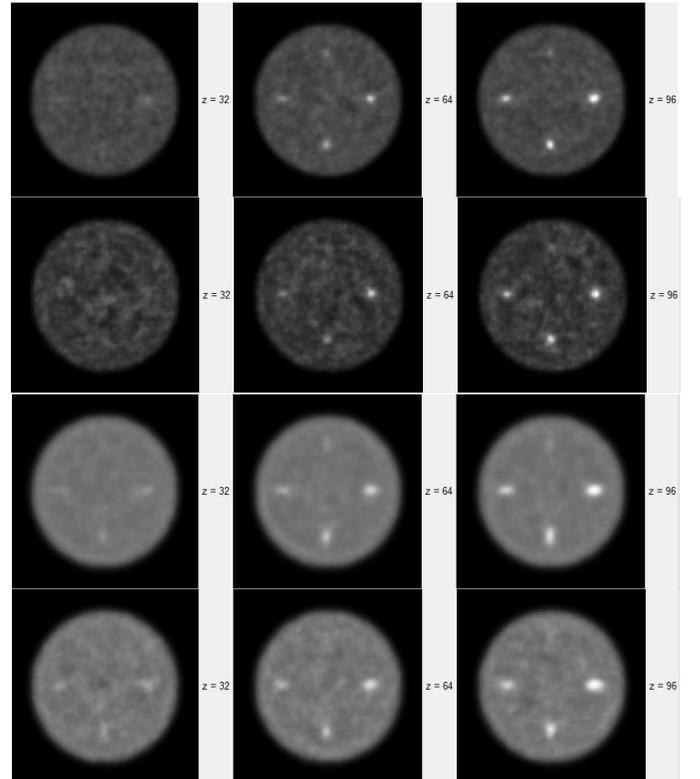

Fig. 1. Example transaxial slices of the reconstructed images showing the hot lesions with contrasts 1:5 (left column), 1:10 (middle

column) and 1:15 (right column). Collimators and scanning times (single detector head) from top to bottom row: HR, 30 min., HR 10 min., HS9, 30 min., HS9, 10 min.

Fig. 2 shows examples of contrast-noise curves for a selected lesion. High resolution collimators do not achieve contrasts as high as the HR, however they offer a better compromise at clinically acceptable noise levels. At the clinical working point the curves for HR 30 min and HS9 10 min scans almost coincide. This indicates that using the HS9 collimator image quality comparable to the HR 30 min scan can be obtained within 10 min, or equivalently with the dose reduced by a factor of 3.

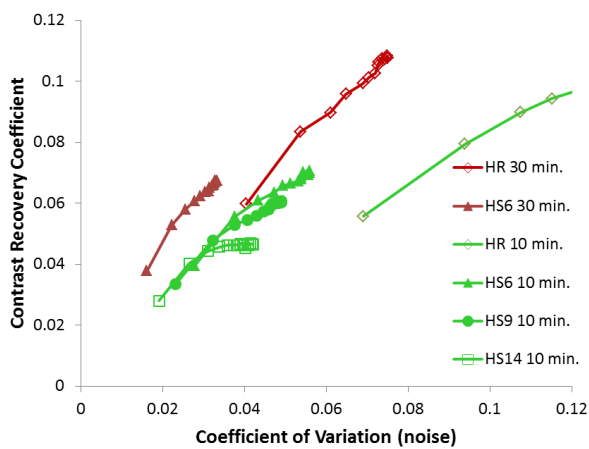

Fig. 2 Contrast-noise curves for the 14 mm 5:1 lesion.

CNR, which provides a rough measure of detectability, is plotted against collimator sensitivity for a few lesions in Fig. 3.

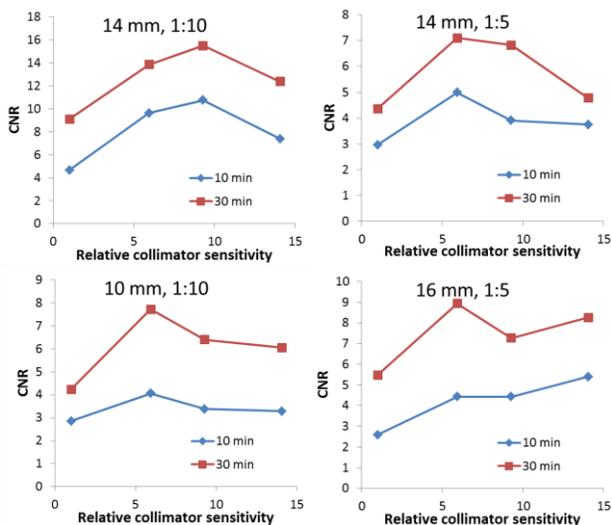

Fig. 3 Contrast-to-noise ratio as a function of collimator sensitivity for selected lesions.

The maximum varies with the size and contrast of the lesion, nevertheless for all lesions it is achieved with a high sensitivity collimator, most frequently the HS6. Similarly as with the contrast noise curves, one can observe that a 10 min. scan obtained with high sensitivity collimators provides CNR comparable to, or higher than a 30 min scan with the HR collimator.

Effective number of visible lesions gives an idea about the system flexibility to image a wide range of lesion sizes and contrasts. Here again the numbers are the highest for high sensitivity collimators.

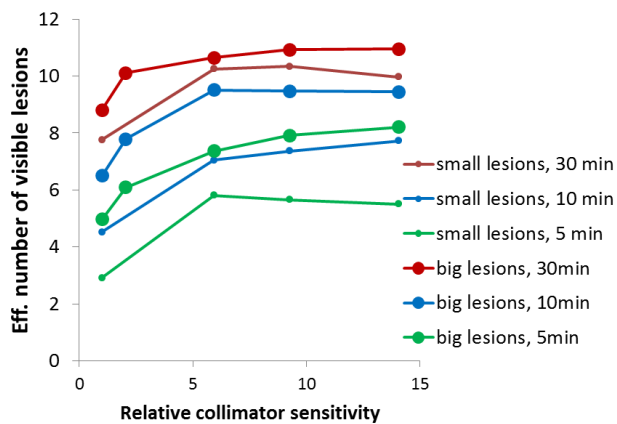

Fig. 4 Effective number of visible lesions (out of 12 in total) as a function of collimator sensitivity for selected lesions.

Human observer studies on detection of larger, lower contrast lesions confirm the quantitative results shown above..

TABLE II. RESULTS OF ROC ANALYSIS OF HUMAN OBSERVER EVALUATION FOR SIGNAL-KNOWN-EXACTLY LESION DETECTION TASK

| Lesion | Collimator Scan time | Performance parameter | | |
|---|---|---|---|---|
| | | AUC | Sensitivity | Specificity |
| 15 mm 1:4 | HR, 30 min. | 0.961 | 86.7 % | 96.7 % |
| | HR, 10 min. | 0.791 | 66.7 % | 86.7 % |
| | HS6, 10 min. | 0.974 | 100 % | 86.7 % |
| | HS6, 30 min. | 1.0 | 100 % | 96.7% |
| 15 mm 1:3 | HR, 30 min. | 0.763 | 73.3 % | 73.3 % |
| | HR, 10 min. | 0.564 | 53.3 % | 56.7 % |
| | HS6, 10 min. | 0.733 | 76.7% | 70% |
| | HS6, 30 min. | 0.968 | 96.7% | 76.7% |

10 minute scan with the high sensitivity collimator yields the sensitivity and specificity comparable to the 30 min. scan with the conventional HR collimator. On the other hand, using high sensitivity collimator in a 30 min. scan improves significantly the detectability of small, low contrast lesions.

IV. CONCLUSIONS

Using different image quality measures, our results show consistently that high sensitivity collimators (6 – 9 times the conventional HR) provide images of superior quality with respect to the task of hot lesions detection on uniform background. With optimal collimators scanning time can be reduced by a factor of 3 or more while maintaining the quality of conventional image. The obtained results are consistent with several other optimisation studies published recently. We are aware that most of the investigations focus on simplified optimisation criteria which may not capture the full complexity of a clinical scan. Nevertheless the results, including those shown in the present paper, consistently indicate that SPECT imaging should move towards high sensitivity collimators and statistical reconstruction with collimator response modelling. More clinical studies are certainly needed to confirm the outcomes of the theory and simulations.

Shortening of the scanning time is equivalent to reducing the activity administered to patient by the same factor. Reducing both of the parameters is important for high throughput and low cost and patient burden screening scans.